\documentclass[aps,pra,reprint,noshowpacs,onecolumn,amssymb]{revtex4}
\usepackage{epsfig}
\usepackage{ifpdf}
\usepackage{graphicx}
\usepackage{amssymb,lineno,mathbbol,upgreek,amsfonts}
\usepackage[tight]{subfigure}
\usepackage{mathrsfs}
\usepackage{bbm}
\usepackage[dvips]{color}
\usepackage{slashed}
\usepackage{natbib}

\usepackage{calligra}
\DeclareMathAlphabet{\mathcalligra}{T1}{calligra}{m}{n}
\DeclareFontShape{T1}{calligra}{m}{n}{<->s*[2.2]callig15}{}

\usepackage{calligra}
\DeclareMathAlphabet{\mathcalligra}{T1}{calligra}{m}{n}
\DeclareFontShape{T1}{calligra}{m}{n}{<->s*[2.2]callig15}{}
\usepackage{enumitem}

\usepackage{amsmath}
\usepackage{enumitem}

\newlength\mylen
\setlist[itemize,1]{leftmargin=*,labelsep=-\mylen}
\begin{document}

\title{Spin-charge separation for paired Dirac fermions in $(1+1)$ dimensions}

\def\correspondingauthor{\footnote{Corresponding author: lhaddad@mines.edu}}

\author{Laith H. Haddad}
\affiliation{Department of Physics, Colorado School of Mines, Golden, CO 80401,USA}
\date{\today}

\begin{abstract}
We study Dirac fermions at finite density coupled to a complex pairing field assumed to obey scalar field theory with quartic self-repulsion. The bulk of our work develops the mathematics that elucidates the propagation of fermionic excitations in such systems as independent spin (boosts) and charge (fermion number) degrees of freedom. A necessary ingredient is the presence of broken $U(1)$ symmetry in the pairing field and decoupling of its density and phase. In the fermion sector, these elements give rise to an emergent spin-dependent gauge coupling which binds in-vacuum spin and charge into elementary fermions, while driving proliferation of unbound spin and charge for finite condensation in the pairing field. Notably, the onset of spin-charge separation is signaled by $\mathcal{P} \mathcal{T}$-symmetry breaking and decoupling of spin components under Lorentz transformations. Our investigation concludes with two theorems that identify generic features of spin-charge separation in such systems. 
\end{abstract}

\maketitle

\section{Introduction}

The phenomenon of spin-charge separation (SCS) is a concept closely associated with dense fermionic quantum systems in reduced dimensions, originally posited and developed within the context of strongly correlated electrons in one-dimensional systems~\cite{Tomonaga1950, Luttinger1963,Haldane1981}. It involves the decoupling of spin and charge excitations, leading to the emergence of distinct quasi-particles carrying either spin or charge but not both. In such systems, quantum fluctuations and strong interactions cause the collective behavior of the fermions to result in separate excitations: spinons, which carry spin but no charge, and chargons, which carry charge but no spin. This separation is a consequence of the breakdown of the standard Fermi liquid picture in one dimension, where fermion-like excitations characterized by a simultaneous presence of spin and charge cease to exist as distinct entities. Though the majority of work on SCS is within condensed matter (lattice) settings, ideal say for investigating the role of SCS in high-temperature superconductivity, e.g., within $t-J$ models of cuprates~\cite{Weng1995,Wen2006}, some have proposed that SCS might be a more general high-energy phenomenon~\cite{Niemi2005,Chernodub2006,Faddeev2007}. The relevance of SCS to relativistic systems beyond one-dimension has been explored with some interesting predictions relevant to our understanding of the standard model of particle physics and its extensions~\cite{Xiong2017}.

In the present work, we study a model of $(1+1)$d relativistic fermions at finite density interacting through both chiral (Gross-Neveu) and complex (difermion) pairing channels. We include additional dynamics for the complex pairing field through a $\phi^4$-type scalar field theory neglecting any back-reaction coming from the fermions. We will show that such models admit a form of SCS -- a decoupling of Lorentz spin and boost structure -- when the difermion channel is attractive enough to form molecules and intermolecular repulsion weak enough to allow for stable $U(1)$ symmetry breaking. Our approach is based on factoring the spinor wavefunction into real and complex parts which contain effects coming from the background density and phase of the pairing field. The complex part further splits into overall and internal phase factors, the former encoding number charge and the latter giving rise to an emergent gauge field that mediates SCS. With the various factors of this decomposition naturally mapping to subgroups of the Lorentz group, the result is a classical dressed spinor which, when elevated to the quantum level, admits a splitting into products of fermionic (for the real factor) and bosonic (for the complex factor) statistics.

Beyond interest in the mathematical nature of our results, we foresee applications to the physics of superconductivity in dense nuclear matter. Indeed, the interplay between the chiral and superconducting channels in dense nuclear systems is a major area of interest in the study of quantum chromodynamics (QCD) at high densities \cite{Deryagin1992,Alford2001_2,Kitazawa2002,Alford2008,Buchoff2010}. Chiral symmetry breaking manifests in the formation of a chiral condensate, altering the nature of quark interactions, while the emergence of superconducting phases involves the formation of Cooper pairs among quarks with different flavors and colors. The intricate balance between these phenomena influences the ground state properties of dense nuclear matter: its equation of state, transport properties, and the potential existence of other exotic phases. Understanding this competition and the potential relevance of SCS promises to shed light on the nature of quantum matter under extreme conditions, providing insights into the behavior of neutron stars, as one example.

 The development of this paper is as follows. In section II, we review some basics of $(1+1)$d Dirac kinematics and dynamics in the presence of pairing fields. Section III illustrates our factorization approach that leads to an SCS interpretation, elucidating momentum scale dependence, group structure, and emergent gauge field. Section IV discusses the landscape of regimes mapped out by SCS. In section IV, we conclude with explicit quasiparticle solutions and general SCS theorems.

\section{Dirac Equation in $(1+1)$ dimensions}

\subsection{Kinematics}

Here, we review kinematics of the Dirac equation.The gamma matrices we use in our (1+1)d model are standard ones. With the time-like signature $g^{\mu \nu}= \mathrm{diag}\left( 1 , - 1 \right)$ and using the $2 \times 2$ gamma matrices defined in terms of the Pauli matrices, we have 
\begin{eqnarray}
\gamma^0 = \sigma_1 \; , \;\;  \gamma^1 = - i  \sigma_2 \;, \;\; \gamma^5 =   \gamma^0   \gamma^1  = \sigma_3 \, ,  \label{myalgebra}
\end{eqnarray}
which satisfy the Dirac algebra  
\begin{eqnarray}
\left\{ \gamma^\mu , \, \gamma^\nu \right\} = 2 g^{\mu \nu}\, . 
\end{eqnarray}
In addition, charge conjugation can be defined as $\psi_C \equiv \gamma^5 \psi^*$, where $\psi_C$ is the charge-conjugated spinor wavefunction determined by the condition $C \gamma^\mu C^{-1} = \pm (\gamma^\mu )^T$, and the Dirac adjoint is defined in the usual way by $\bar{\psi} = \psi^\dagger \gamma^0$. Varying the free-particle action associated with the Lagrangian density 
\begin{eqnarray}
\hspace{-1pc} \mathcal{L}  &=&  \bar{\psi}\left( i \gamma^\mu  \partial_\mu - m  + \mu \gamma^0  \right)   \psi \, ,   \label{Lag}
\end{eqnarray}
with respect to $\bar{\psi}$, gives the Dirac equation for two-dimensional spinor solutions $\psi = (\psi_1, \psi_2)^T$
\begin{eqnarray}
i (\partial_t - \partial_x ) \psi_1  - m \psi_2 + \mu \psi_1 &=& 0  \, ,\\
i (\partial_t + \partial_x ) \psi_2  - m \psi_1 + \mu \psi_2 &=& 0 \, . 
\end{eqnarray}
This set of coupled equations has plane-wave solutions
\begin{eqnarray}
\psi(x, t) =   e^{i (px  -  E t)}  \left( \begin{array}{l}
 \sqrt{ \frac{ E  + \mu - p  }{m} }  \\
           \sqrt{ \frac{m}{ E  + \mu - p  }}       \end{array} \right) \, . 
\end{eqnarray}
The associated single-particle momentum space Dirac operator for finite mass and chemical potential is
\begin{eqnarray}
\mathcal{D} &=& \left( \begin{array}{ll}
  E + \mu + p      &   \;\;\;\; - m    \vspace{0pc}\\
\; \;  -m  &       E + \mu - p          \end{array} \right) \, , 
\end{eqnarray} 
which gives the Dirac equation
\begin{eqnarray}
\mathcal{D} \,  \psi = 0 \, ,  \label{DopEq}
\end{eqnarray} 
 for spinor solutions
 \begin{eqnarray}
\psi  &=& \left( \begin{array}{ll}
  \psi_1     \\
  \psi_2       \end{array} \right) \, . 
\end{eqnarray} 
The Dirac operator can be reparametrized to the hyperbolic form 
\begin{eqnarray}
\mathcal{D} &=& \left( \begin{array}{ll}
 \; e^\eta      &   \;  - 1   \vspace{0pc}\\
  -1 &     \;    e^{-\eta}       \end{array} \right) \, , 
\end{eqnarray} 
where $\mathrm{cosh}\, \eta \equiv (E + \mu)/m$, $\mathrm{sinh} \, \eta \equiv  p/m$, and $\mathrm{tanh} \, \eta = p/(E+ \mu)$. Here, $\eta$ is the usual rapidity associated with generators of the $(1+1)$d Lorentz transformations $\Lambda$
\begin{eqnarray}
S\left[ \Lambda \right] &=& \left( \begin{array}{ll}
 \; e^\eta      &   \;   0   \vspace{0pc}\\
   \; 0  &      e^{-\eta}       \end{array} \right)  = e^{\eta \gamma^5} \, .
\end{eqnarray} 
Thus, the relativistic limit of the in-medium (finite $\mu$, small $m$, large Fermi surface) Dirac operator amounts to an ordinary Lorentz boost through elements of the hyperbolic subgroup of $SL(2, \mathbb{R})$. It is straightforward to show that solutions of Eq.~(\ref{DopEq}) have the form 
\begin{eqnarray}
\psi  &=& \phi \left( \begin{array}{l}
  e^{- \eta/2}    \\
  \,  e^{\eta/2}     \end{array} \right) \, .  \label{SinglePartState}
\end{eqnarray} 
Alternatively, one could choose a parametrization such that $\mathrm{cos}\, \phi \equiv m/(E + \mu)$, $\mathrm{sin} \, \phi \equiv  p/(E + \mu)$, and $\mathrm{tan} \, \phi  = p/m$, which gives 
\begin{eqnarray}
\mathcal{D} &=& \left( \begin{array}{ll}
 1 + \mathrm{sin} \, \phi      &   \;  -  \mathrm{cos} \, \phi  \vspace{0pc}\\
  - \mathrm{cos} \, \phi &     \;     1 - \mathrm{sin} \, \phi        \end{array} \right) \, , 
\end{eqnarray} 
which has solutions of the form 
\begin{eqnarray}
\psi  &=& \theta  \left( \begin{array}{l}
  \mathrm{cos} \, \phi/2   \\
    \mathrm{sin} \, \phi/2    \end{array} \right) \, . 
  \end{eqnarray}

One may also obtain a complex form by choosing the reduced gamma matrices to be
\begin{eqnarray}
\gamma^0 =  - \sigma_1 \; , \;\;  \gamma^1 = i  \sigma_3 \;, \;\; \gamma^5 =   \gamma^0   \gamma^1  =  -  \sigma_2 \, , \;\; C = \gamma^0 \,  .  \label{altalgebra}
\end{eqnarray}
With this choice the (linear) Dirac equation is 
\begin{eqnarray}
( m - \partial_x  ) \psi_1  + (i \partial_t + \mu) \psi_2  &=& 0  \, ,\\
  (i \partial_t + \mu) \psi_2   + (m +  \partial_x   ) \psi_1 &=& 0 \, , 
\end{eqnarray}
which has the plane-wave solutions
\begin{eqnarray}
\psi(x, t) =   e^{i (px  -  E t)}  \left( \begin{array}{l}
 \sqrt{ \frac{ E  + \mu   }{m - i p} }  \\
           \sqrt{ \frac{m - i p }{ E  + \mu  }}       \end{array} \right) \, . 
\end{eqnarray}
The momentum space Dirac operator now takes the form
\begin{eqnarray}
\mathcal{D} &=& \left( \begin{array}{ll}
 -( m -  i  p )      &   \;\;    \; E - \mu \vspace{0pc}\\
 \;\; E- \mu  &     \;\; - ( m +  i  p )          \end{array} \right) \, , 
\end{eqnarray} 
which can be reparamtrized to the trigonometric form 
\begin{eqnarray}
\mathcal{D} &=& \left( \begin{array}{ll}
 \; e^{-i \phi}      &   \;  - 1   \vspace{0pc}\\
  -1 &     \;    e^{i \phi }       \end{array} \right) \, , 
\end{eqnarray} 
where $\mathrm{cos}(\phi) \equiv  m/(E -  \mu)$, $\mathrm{sin} ( \phi ) \equiv  p/(E - \mu)$, and $\mathrm{tan}( \phi )= p/m$. Solutions then have the form 
\begin{eqnarray}
\psi  &=& \theta \left( \begin{array}{l}
  e^{ i \phi /2}    \\
  \,  e^{  -  i \phi/2}     \end{array} \right) \, .  \label{SinglePartState}
\end{eqnarray} 
Note that, in this complex form, Lorentz transformations are now given by
\begin{eqnarray}
S\left[ \Lambda  \right] &=& \left( \begin{array}{ll}
 \; e^{- i \phi}     &   \;   0   \vspace{0pc}\\
   \; 0  &      e^{i \phi}       \end{array} \right)  = e^{-  \phi \gamma^0 \gamma^5} \, .
\end{eqnarray}

\subsection{Pairing fields}

In $(1+1)$ dimensions the interactions that give rise to pairing fields are those associated with scalar, pseudoscalar, vector, axial vector, difermion, and density interactions which can be expanded, respectively, as
\begin{eqnarray}
(\bar{\psi } \psi )^2 &=& 2 \left[ \cos( 2 \beta) + 1 \right] |\psi_1|^2 | \psi_2|^2  \, , \\
(\bar{\psi } \gamma^5  \psi )^2 &=& 2  \left[ \cos( 2 \beta) - 1 \right] |\psi_1|^2 | \psi_2|^2  \, , \\
(\bar{\psi } \gamma^\mu  \psi )^2 &=& 4 |\psi_1|^2 | \psi_2|^2  \, ,  \\
(\bar{\psi } \gamma^5 \gamma^\mu  \psi )^2 &=& -  4 |\psi_1|^2 | \psi_2|^2  \, , \\
| \psi^T   C \psi |^2 &=&    -  2   \cos( 2 \beta)   |\psi_1|^2 | \psi_2|^2     +    |\psi_1|^4 +   |\psi_2|^4    \, , \\
( \bar{\psi } \gamma^0  \psi )^2 &=&     2    |\psi_1|^2 | \psi_2|^2     +    |\psi_1|^4 +   |\psi_2|^4    \, , 
\end{eqnarray}
where $\beta$ is the relative phase between the upper and lower spin components. This phase is superfluous to spinors in $(1+1)$d (the overall phase is not) and emerges naturally out of the complex difermion field. If we set $\beta = 0$, for instance, the interaction types reduce considerably to 
\begin{eqnarray}
(\bar{\psi } \psi )^2 &=&     (\bar{\psi } \gamma^\mu  \psi )^2  =  -  (\bar{\psi } \gamma^5 \gamma^\mu  \psi )^2    =   4 |\psi_1|^2 | \psi_2|^2       \, , \\
| \psi^T   C \psi |^2 &=&    (  |\psi_1|^2   -    |\psi_2|^2 )^2    \, , \\
( \bar{\psi } \gamma^0  \psi )^2 &=&     (   |\psi_1|^2 +   |\psi_2|^2 )^2    \, ,  \\
(\bar{\psi } \gamma^5  \psi )^2 &=&   0  \, ,
\end{eqnarray}
from which, without loss of generality, we may focus on the following pairing fields
\begin{eqnarray}
\bar{\psi } \psi    &=&      2   \, |\psi_1|  | \psi_2|       \, , \\
 \psi^T   C \psi   &=&     \psi_1^2   -    \psi_2^2    \, , \\
 \bar{\psi } \gamma^0  \psi  &=&     |\psi_1|^2 +   |\psi_2|^2   \, . 
\end{eqnarray}

\subsection{Dynamics induced by scalar and difermion channels}

Incorporating these three pairing fields, our basic starting point will be the Lagrangian density for relativistic fermions with finite mass and chemical potential interacting through scalar meson and superconducting channels given by 
\begin{eqnarray}
\hspace{-1pc} \mathcal{L}_\psi  &=&  \bar{\psi}\left( i \gamma^\mu  \partial_\mu - m  + \mu \gamma^0  \right)   \psi  + \frac{g_s}{2}  \left(   \bar{\psi}     \psi  \right)^2    + \frac{g_d}{2}  \left(   \bar{\psi}  C   \bar{\psi}^T  \right) \! \left(   \psi^{T}   C     \psi      \right) . \label{FundLag}
\end{eqnarray}
The physical parameters $m$, $\mu$, $g_s$ and $g_d$ are the mass, chemical potential, scalar and difermion couplings, respectively, and $C$ is the charge conjugation operator. In addition, we might consider somewhat separately the dynamical equations for both real chiral and complex difermion fields 
\begin{eqnarray}
 \mathcal{L}_\sigma    =       \frac{1}{2}   \partial_\mu \sigma \partial^\mu \sigma  - \frac{1}{2} m_\sigma^2 \sigma^2  -  \frac{g_{\sigma \Delta}}{2 !  }  \, \sigma^2  |\Delta|^2   \; , \;\;\;\; \;\;\; \mathcal{L}_\Delta  =   \frac{1}{2}   \partial_\mu \Delta^* \partial^\mu \Delta  - \frac{1}{2} m_\Delta^2 |\Delta|^2  -  \frac{g_\Delta}{4 !  }  \, |\Delta|^4 \, .   \label{NLKG}
 \end{eqnarray}
 In particular, the equations of motion for the difermion density and phase are 
 \begin{eqnarray}
 \partial_\mu \partial^\mu \! \rho_\Delta + m_\Delta^2 \rho_\Delta  + \frac{g_\Delta}{6} \rho_\Delta^3 = \rho_\Delta  \partial_\mu \beta_\Delta \partial^\mu \beta_\Delta \, , \;\;  \partial_\mu  \! \left(  \rho_\Delta  \partial^\mu \beta_\Delta \right) = 0  \, . \label{densityphasedec}
 \end{eqnarray}
 with the conserved current in the right equation acting as a source for density fluctuations in the left one. The low-temperature mean-field limit of Eq.~(\ref{FundLag}) allows for several non-vanishing condensates
\begin{eqnarray}
\rho = \langle \bar{\psi } \gamma^0 \psi \rangle \, , \;\;\; \sigma = \frac{g_s}{2}  \langle \bar{\psi} \psi \rangle \, , \;\;\;  \Delta =   \frac{g_d}{2} \langle  \psi^T C \psi \rangle   \, , \;\;\;  \bar{\Delta}   =   \frac{g_d}{2}  \langle  \bar{\psi} C \bar{\psi}^T \rangle = - \Delta^* \, , 
\end{eqnarray}
which are respectively, the average density, chiral, difermion and conjugate difermion condensates. In the mean-field approximation, Eq.~(\ref{FundLag}) leads to the modified Dirac equation
\begin{eqnarray}
i \left[ \gamma^0\!  \left(\partial_t - i \mu \right) - \gamma^1\!  \left(\partial_x + i \bar{\Delta}   \right) \right] \psi - \left( m - \sigma \right) \psi = 0 \, . \label{ModDirac}
\end{eqnarray}
It is important to recall that mean-field descriptions are justified at low temperatures and weak interactions or if one assume a standard large-$N$ limit description. To obtain plane wave solutions of Eq.~(\ref{ModDirac}) one could solve the $4 \times 4$ system for the Fourier components 
\begin{eqnarray}
 \left( \begin{array}{l l l l }
      \;   \sigma - m  \; &\; \;\;   \; 0  \;  & \; E_- + Re \Delta \; & \; \;\; Im \Delta \;  \\
            \; \;\;  \;  0 \;    &  \; \sigma - m \;  & \; \; - Im \Delta \;  & \; E_- + Re \Delta  \;    \\
           \;   E_+ - Re \Delta   \;   &\;  - Im \Delta \;  & \; \sigma - m \;  &  \;\;\; \;\; 0 \;   \\
              \; \;\; Im \Delta \;       &  \;  E_+ - Re \Delta   \;     & \;\;\; 0 \;  &   \; \sigma - m \; 
            \end{array} \right)
             \left( \begin{array}{l}
        Re  \,  \tilde{\psi}_1  \\
         Im \,  \tilde{\psi}_1 \\
         Re   \,  \tilde{\psi}_2 \\
           Im  \,   \tilde{\psi}_2
            \end{array} \right)   =  0  \, , 
  \end{eqnarray}
from which one obtains the dispersion through the condition
\begin{eqnarray}
&&(\sigma -m )^2 - 2 (\sigma -m )^2 Im \Delta^2  + (  E_- + Re \Delta)^2 Im \Delta^2 +    Im \Delta^4 \\
&&- 2 (\sigma -m )^2  (  E_- + Re \Delta)  (  E_+ - Re \Delta) +   (  E_- + Re \Delta)^2  (  E_+ - Re \Delta)^2  +    Im\Delta^2     (  E_+ - Re \Delta)^2 = 0  \, . 
\end{eqnarray}
The Fourier components, generally complex, are then given by 
\begin{eqnarray}
\tilde{\psi}_1   &=&   \frac{- (\sigma -m )^2   (E_- + Re \Delta)  +   (E_- + Re \Delta)^2  (E_+ - Re \Delta)  + Im \Delta^2   (E_+ - Re \Delta) }{   (\sigma -m ) \left[ (\sigma -m )^2    -    Im \Delta^2   -  (  E_- + Re \Delta) (  E_+ - Re \Delta)   \right]                } \\
                    &+&    i   \,  Im\Delta  \,  \frac{(\sigma -m )^2    -   (E_+ -  Re \Delta)^2    -  Im \Delta^2 }{   (\sigma -m ) \left[ (\sigma -m )^2    -    Im \Delta^2   -  (  E_- + Re \Delta) (  E_+ - Re \Delta)   \right]                } \, , \\
\tilde{\psi}_2   &=&   1 +    i   \, Im \Delta \,   \frac{  (E_-  +  Re \Delta)     -  (E_+  -  Re \Delta)     }{   \left[ (\sigma -m )^2    -    Im \Delta^2   -  (  E_- + Re \Delta) (  E_+ - Re \Delta)   \right]                } \, . 
                    \end{eqnarray}
Instead of taking this approach, one may choose a solution of the form $\xi \psi$ where $\psi$ satisfies the vacuum Dirac equation and $\xi$ contains background effects. For a uniform background (London limit), the modified plane-wave solutions have the form 
\begin{eqnarray}
\psi(x, t)   =   e^{i \left( p x - E t \right) }    \left( \begin{array}{l}
     e^{- \eta'/2}  \\
       e^{\eta'/2}   \end{array} \right) \, \label{modsol}
  \end{eqnarray}
with the modified boost parameters given by
\begin{eqnarray}
  e^{\eta' }  \equiv \sqrt{\frac{E + \mu + (p + \bar{\Delta})}{E + \mu - (p +  \bar{\Delta })} }  =    \sqrt{\frac{E_+ }{E_-}}  \,     \sqrt{\frac{1   +  \bar{\Delta }/E_+ }{1 -  \bar{\Delta} / E_-}}  = e^\eta \, e^\zeta  \, , 
      \end{eqnarray}
and where we have defined $E_\pm = E + \mu \pm p$. Dependence on the difermion field $\Delta$ resides in $\zeta$, with $\eta$ depending purely on kinematic variables. As a consistency check it is instructive to express the parametrization of single-particle states in Eq.~(\ref{SinglePartState}) purely in terms of the condensate parameters (no kinematic variables) by first expressing the latter as 
\begin{eqnarray}
\rho =  2 |\phi|^2 \,  \mathrm{cosh}  ( \zeta')  \, , \;\;\; \sigma =  2 |\phi|^2  \, , \;\;\;  \Delta =  - 2 \phi^2  \, \mathrm{sinh}  ( \zeta')   \, , \;\;\;  \bar{\Delta}   =     2 {\phi^*}^2 \,  \mathrm{sinh}  ( \zeta')  \, , 
\end{eqnarray}
then inverting. Without loss of generality, we may take $\phi \to  i \phi $, ($\phi , \Delta \in \mathbb{R}$), whereby we obtain 
\begin{eqnarray}
e^{\zeta'}     =  \frac{  \rho + \bar{\Delta} }{\sigma} =   \pm     \sqrt{ \frac{\rho + \bar{\Delta}}{\rho -  \bar{\Delta} }            }     =       \pm     \sqrt{ \frac{1 +  \bar{\Delta}/\rho }{ 1 -  \bar{ \Delta} /\rho }             }        \; ,  \;\;\; \phi = \pm  \sqrt{\frac{\sigma}{2}} \; , \;\;\; \rho^2 - \bar{\Delta}^2 = \sigma^2 \,.  \label{InMedCond}
\end{eqnarray}
We see that in the limit $p \to 0$, $\zeta \to \zeta'$ since here $E_+ = E_- \sim \rho$.

\section{Spin-charge separation}

The basic premise is to consider a decomposition for classical Dirac solutions which one may naturally interpret as separation of spin and charge degrees of freedom. Let us continue along the lines of Eq.~(\ref{modsol}) by considering a field theory expansion with noninteracting fields as basic elements but dressed by the background fields. The modified boost parameters may then be separated into factors of the form
\begin{eqnarray}
 e^{\eta' } &\equiv&  \sqrt{\frac{E + \mu + (p + \bar{\Delta})}{E + \mu - (p +  \bar{\Delta})}}  \\
  &=&    \sqrt{\frac{E_+ }{E_-}}  \,     \sqrt{\frac{1   +   Re \bar{\Delta}/E_+ }{1 -    Re \bar{\Delta} / E_-}}  \,  \sqrt[4]{\frac{1   +     Im \bar{ \Delta}^2/(E_+ +  Re  \bar{\Delta})^2 }{1 +  Im  \bar{\Delta}^2 / (E_- -   Re  \bar{\Delta})^2   }}  \, e^{ i \left\{  \mathrm{tan}^{-1}\left[  Im \bar{\Delta}/(E_+ +     Re \bar{\Delta}) \right]    -         \mathrm{tan}^{-1}\left[  Im \bar{\Delta}/(E_- -   Re \bar{\Delta}) \right]                              \right\} /2  }   \label{dressed1}  \\
   &\equiv& e^\eta \, e^\zeta  \,  \phi \, e^{i \beta} \, , 
   \end{eqnarray}
   where factors in the last line are identified in order with those in the previous step, and where the kinematic variables are encapsulated in $E_\pm = E + \mu \pm p \in \mathbb{R}$ and the difermion field is generally complex, $\bar{\Delta} \in \mathbb{C}$. Space-time dependent fields then acquire the form 
   \begin{eqnarray}
       \psi_\pm'   =    e^{  \mp \zeta  /2  }   \,  e^{  \mp i \beta /2  } \,  \phi^{\mp 1/2}  \,   \psi_\pm \, , 
      \end{eqnarray}
with the primed (unprimed) field on the left (right) indicating background dressed (elementary) fields. At the quantum level, one may combine the last three factors to form a composite bosonic charge field $\phi_\pm  \equiv \phi^{\mp 1/2}   \, e^{  \mp  i \beta /2  }  \, \psi_\pm $, such that  
 \begin{eqnarray}
       \psi_\pm'   =    e^{  \mp \zeta /2  } \,  \phi_\pm \, ,   \label{basictrans}
      \end{eqnarray}
where now the exponential prefactor must obey fermionic statistics so that the dressed field retains the correct statistics. This line of reasoning which suggests splitting into fermionic and bosonic parts in the presence of a nontrivial background requires a bit more justification. Let us assume anticommutation rules for dressed fields with spin indices $a, \, b$ at positions $x, \, y$. This implies the following for the split fields
 \begin{eqnarray}
 \left\{  \psi_ a'(x) , \,  \psi_b'^*(y)     \right\}     &= &    e^{ \zeta_a(x)/2} \phi_a(x) \,    e^{\zeta_b(y)/2} \phi_b^*(y)   +   e^{ \zeta_b(y)/2} \phi_b^*(y)  \,   e^{ \zeta_a(x)/2} \phi_a(x)   \, \\
                                                                      &=&      e^{ \zeta_a(x)/2}   e^{\zeta_b(y)/2}   \left\{    \phi_a(x)   \phi_b^*(y)     -     \phi_b^*(y)    \phi_a(x)      \right\}  +     \delta_{a  b}( x- y) \, \phi_b^*(y)    \phi_a(x)  \,\\
                                                                      &=&     \delta_{a  b}( x- y) \, \left\{    e^{ \zeta_a(x)/2}   e^{\zeta_b(y)/2}        +     \phi_b^*(y)    \phi_a(x)     \right\}  \,. 
                                                                                                                                            \end{eqnarray}
  Thus, assuming fermionic anticommutation relations for the exponential factors, which are essentially coherent states of the bosonic field $\zeta$, forces the complex factors to satisfy bosonic commutation rules. The presence of the term inside the brackets in the last line indicates that a fundamental  requirement for this description to work is the presence of background density.

   \subsection{Momentum-scale dependence}
   \label{MomScale}
   
   The emergence of SCS is momentum dependent as can be seen from Eq.~(\ref{dressed1}). Let us examine the form of Eq.~(\ref{dressed1}) in the deep material region where $\mu$ and $\bar{\Delta}$ are large and $m$ is small. We consider the two limits of large versus small single-particle momenta. For large $p$ and keeping only lowest nonzero contributions, $E + \mu \to p \, \Rightarrow \,   E_- \to 0$ and $E_+ \gg |\bar{\Delta}| \, \Rightarrow \,  \beta \to \mathrm{tan}^{-1} \! \left( Im \bar{\Delta}/Re \bar{\Delta}\right) \!/2 = \mathrm{arg}\! \left(\bar{\Delta} \right)\!/2$ and 
   \begin{eqnarray}
   \psi_\pm' \to       \left(\frac{\bar{\Delta}}{E_-} \right)^{\! \! \pm 1/2} \! \! \! \psi_\pm \,  =   \sqrt{\rho} \, e^{i \theta_N} \!  \left(\frac{\bar{\Delta}}{E_+} \right)^{\! \! \pm 1/2}  . \label{largeP}
      \end{eqnarray}
In this regime the phase of $\bar{\Delta}$ amounts to a global $U(1)$ transformation of the elementary spinor field since it no longer depends on $p$. The magnitude $|\bar{\Delta}|$ replaces the factor of $E_-$ that appears in the numerator (denominator) of the upper (lower) component of spin acting to counter large boosts in the elementary spinor field. Thus, at large momentum we recover the original vacuum spin structure but with the density $|\bar{\Delta}|$ in the role of particle mass $m$. In this limit the current associated with the internal phase vanishes leaving only propagating bound spin and fermion number charge. 

In contrast, for small $p$, to leading order we find that 
    \begin{eqnarray}
   \psi_\pm' \to        \sqrt{\rho}    \left(  \frac{1+ Re \bar{\Delta}/\rho            }{ 1-  Re \bar{\Delta}/ \rho } \right)^{\! \! \mp 1/2}  \!\! \! e^{  \mp  i   \beta_\Delta/2  } \,   e^{ i  \theta_N } \,  \simeq  \,       \left(  \frac{1+ d_1 q_{\Delta}/\rho_0            }{ 1-  d_1  q_{\Delta}/ \rho_0 } \right)^{\! \! \mp 1/2}  \!\! \! e^{  \mp  i  a_1 q_\beta  /2  } \,   \sqrt{\rho_0  } \, e^{ i  c_1 p }               , \label{chargprop}
      \end{eqnarray}   
   where we have expanded the difermion density and phase, $\bar{\Delta}$ and $\beta_\Delta$, and the charge density $\rho$ and phase $\theta_N$ in terms of their respective momenta: $| \bar{\Delta}| \simeq d_1 q_\Delta + \dots , \;     \beta_{\Delta} \simeq a_1 q_\beta + \dots  , \; \rho \simeq \rho_0 + \dots , \;  \theta_N \simeq c_1 p + \dots  \; $. It is interesting that by including fluctuations in $\bar{\Delta}$ with momentum $q_\Delta$ the prefactor in Eq.~(\ref{chargprop}) takes the form of a propagating spin degree of freedom. The density $\rho$ and phase $\theta_N$ together define charge propagation at momentum $p$. There is the appearance of an additional (internal) phase that propagates with characteristic momentum $q_\beta$ (we will identify this as a gauge potential). Generally then at the two extreme limits that contrast phase and background density momenta, i.e., $p,   q_\beta  \gg q_\Delta$ and $p   ,  q_\beta     \ll q_\Delta$, the system will be described by elementary spinors propagating over a background in the former case, versus decomposed spin and charge plus an additional gauge structure propagating independently in the latter case:
   \begin{eqnarray}
  \textrm{ spinor + background}    \;\;\;  \leftrightarrow \;\;\;  \textrm{spin + gauge + charge }   \, . 
   \end{eqnarray}
   Note that large fluctuations in phase describes strong depletion of the difermion condensate and dissociation of difermion bound states, in contrast to the case of pure difermion density fluctuations.

 \subsection{Group structure}
\label{groupstructure}

 We can further characterize the splitting into spin and charge degrees of freedom by contrasting the group structures in the elementary fermion and spin-charge formulations. First note that Eq.~(\ref{ModDirac}) is symmetric under parity when parity transformations are defined in the usual way through $\mathcal{P}: \psi(x, t)  \rightarrow  \gamma^0  \psi(-x, t)$, since $\bar{\Delta}$ is odd and $\sigma$ and $m$ are even under $\mathcal{P}$. Invariance of Eq.~(\ref{ModDirac}) under time-reversal however, i.e., $\mathcal{T}: \psi(x, t)  \rightarrow  \gamma^0  \psi^*(x, - t)$, holds only if $\bar{\Delta }= - \bar{\Delta}^* \Rightarrow Re\bar{\Delta} = 0$, in the chiral limit $m, \sigma =0$. Recall that in $(1+1)$d our choice for the gamma matrices forces us to work with an anti-Hermitian matrix, $\gamma^1 \in SO^+(2) < SL^+(2, \mathbb{R})$, yet we are able to render it Hermitian through the action of $\gamma^0 \in SO^-(2) < SL^{-}(2, \mathbb{R})$. Since we only have one spatial gamma matrix to contend with in the present dimensionally reduced problem, we may attempt to define parity inversion using $\gamma^1$ instead of $\gamma^0$, i.e., $\mathcal{P}: \psi(x, t)  \rightarrow  \gamma^1  \psi(-x, t)$. One finds that under this definition of parity, $\bar{\Delta}$ and $\sigma$ are both odd so that Eq.~(\ref{ModDirac}) is invariant when the fermion mass is purely generated by the scalar condensate.

 We can identify the total charge field in Eq.~(\ref{dressed1}) as the two-component composite object whose elements decompose as a product of group factors acting on the superposition of basis vectors $n_1 = (1 , \, 0)^T$, $n_2 = (0, \, 1 )^T \to  \Phi = g \, (n_1 + n_2)^T/\sqrt{2}$, where $g \in U(1) \times SL(2, \mathbb{C})$ and $\Phi \cong U(2)$. Explicitly, 
   \begin{eqnarray}
 \underbrace{  \Phi  }_\text{charge field}    =   \underbrace{  e^{i \theta}  }_\text{$U(1)_N$}  \;  \underbrace{    \left( \begin{array}{ll} 
  e^{-\eta /2}     &   \;  0   \vspace{0pc}\\ 
  0 &     \;    e^{\eta/2}       \end{array} \right)  }_\text{ $SL(2, \mathbb{R})_\mathrm{boost}$  } \;   \underbrace{    \left( \begin{array}{ll} 
 \phi^{1/2}      &   \;  \;  0   \vspace{0pc}\\ 
  0  &     \;    1/\phi^{1/2}      \end{array} \right)  }_\text{$SL(2, \mathbb{R})_\Delta $} \;  \underbrace{    \left( \begin{array}{ll} 
 e^{- i \beta/2}      &   \;  0   \vspace{0pc}\\ 
\;   0 &     \;    e^{i \beta/2 }       \end{array} \right)  }_\text{$SU(2)_\mathrm{\Delta}$}  \;   \underbrace{     \frac{1}{\sqrt{2}}   \left( \begin{array}{l}                
                    1    \\ 
                     1 
   \end{array}    \right) }_\text{unit vector}     \equiv      \left( \begin{array}{l}                   
                    \phi_1   \\ 
                    \phi_2
   \end{array}    \right)         \label{group1} \,  .
       \end{eqnarray}
The group factors on the right are associated, respectively, with fermion number charge $N$ and the remaining three combine to form a representation of the hyperbolic ($A \cong \mathbb{C}_{> 0}$) subgroup $SL(2, \mathbb{R})_\mathrm{boost}  \times  SL(2, \mathbb{R})_\Delta \times  SU(2)_\mathrm{\Delta} < SL(2, \mathbb{C})_{\mathrm{boost} + \Delta}$, which follows from the standard Iwasawa $K \cdot A \cdot N$ decomposition. The physical distinction between boosts of elementary spinors and background effects induced by the difermion field naturally decompose the hyperbolic subgroup into a product of associated elements: two subgroups of $SL(2, \mathbb{R})$ and one of $SU(2)$. We choose to absorb the boost matrix for the elementary spinors into $\Phi$ since we are interested in the deep material regime in Eq.~(\ref{chargprop}) where vacuum spinors are no longer appropriate descriptions. The spin field in this regime is then just the remaining factor in Eq.~(\ref{dressed1}) associated with the background driven quasi-boosts in Eq.~(\ref{chargprop}). We can express it in terms of group elements as $\chi = a \, (n_1 + n_2)^T/\sqrt{2}$, where $a \in SL(2, \mathbb{R})$:
\begin{eqnarray}
 \underbrace{  \chi  }_\text{spin field}    =    \underbrace{    \left( \begin{array}{ll} 
  e^{-\zeta /2}     &   \;  0   \vspace{0pc}\\ 
  0 &     \;    e^{\zeta/2}       \end{array} \right)  }_\text{ $SL(2, \mathbb{R})_{\Delta-\mathrm{boost}}$  } \;  \underbrace{     \frac{1}{\sqrt{2}}   \left( \begin{array}{l}                
                    1    \\ 
                     1 
   \end{array}    \right) }_\text{unit vector}     \equiv      \left( \begin{array}{l}                   
                    \chi_1   \\ 
                    \chi_2
   \end{array}    \right) \, .     \end{eqnarray} 
The full in-medium wavefunction is then expressed as
\begin{eqnarray}
\psi = \chi \Phi  \, =  \, e^{ - \frac{1}{2} \zeta \gamma^5}  \Phi  \, . 
\end{eqnarray}

Returning to the concept of parity, it is significant that $\gamma^1$ is anti-idempotent under multiplication and the group that it generates $\mathcal{G} = \left\{ \mathbb{1}, \, \gamma^1, - \mathbb{1}, \, - \gamma^1     \right\}$ is isomorphic to the fourth roots of unity $\mathcal{U} = \left\{ i^0, \, i^1 , \, i^2 , \, i^3    \right\}$ generated by $e^{i \pi/2}$ under multiplication equivalent to $\mathbb{Z}_4$, the cyclic group of order four. The definition of parity that uses $\gamma^1$ can be interpreted as a rotation by $180^\mathrm{o}$ in 2D about the y axis by noting that 
\begin{eqnarray}
\gamma^1  \,  =  \, \underbrace{    \left( \begin{array}{ll} 
   0  &   \; -    1  \vspace{0pc}\\  
   1 &     \; \; \;  0            \end{array} \right)  }_\text{1D spin parity }  \,  =  \,  \underbrace{    \left( \begin{array}{ll} 
  \mathrm{cos}(\phi /2) &   \; - \mathrm{sin}(\phi/2)  \vspace{0pc}\\ 
 \mathrm{sin}(\phi/2) &     \; \; \; \mathrm{cos}(\phi /2)\end{array} \right)_{\! \phi = \pi}  }_\text{2D spin rotation $\cong K < SL(2, \mathbb{R})$  }          =  \; e^{- i (\phi/2 ) \sigma_y } \vert_{\phi = \pi }  \; \; .
 \end{eqnarray}
Here $\phi$ is the polar angle that parametrizes spin rotations in the $x, z$-plane which are isomorphic to the $K$ subgroup in the Iwasawa decomposition. Such rotations map to parity inversion when restricted to $\phi = \pi$. Note that the other gamma matrix $\gamma^0$ offers no such interpretation. Thus, even though there are no rotations in 1D, two parity inversions defined in this way result in the characteristic overall factor of $-1$. Since the part of Eq.~(\ref{group1}) parameterized by $\eta$ solves the vacuum Dirac equation, it must transform in the same way under parity acquiring an overall factor of $-1$ by itself. We have seen that parity inversion in Eq.~(\ref{dressed1}) amounts to flipping the signs of both $\bar{\Delta}$ and $p$, and one may expand $\bar{\Delta} \simeq   d_1 q_\Delta + d_3   q_\Delta^3 + \dots$ in the difermion momentum $q_\Delta$ in Eq.~(\ref{chargprop}) to recover a (quasi) spin structure when $p \to 0$. In this way, a hand off of spin from the elementary fermions to the background fluctuations takes place through a crossover region (as discussed in Sec.~\ref{MomScale}).

  Having identified $\chi$ as the carrier of spin, one may consider a Maxwellian $U(1)_M$ gauge transformation that sends $\psi \to \exp(i \theta_M) \,  \psi$. Since $\chi \in SL(2 ,\mathbb{R})$, it cannot change under such gauge transformations but $\Phi \in U(2)$ has room within its $U(1)_N$ factor, hence $\Phi$ is the natural carrier of charge. An additional internal symmetry can be identified in the decomposition $\chi \Phi$ if one treats $\chi$ as an embedding $SL(2, \mathbb{R})  \hookrightarrow SL(2, \mathbb{C}) \cong SL(2, \mathbb{R}) \times SU(2)$ by incorporating half of the internal phase $\beta$ into $\chi$. A local internal $SU(2)$ symmetry can then be identified that sends $\chi \to \chi \, g$ and $\Phi \to g^{-1} \Phi$ with $g \in SU(2)$. This internal symmetry is then broken by the presence of local fluctuations in $\beta$ through phase (versus density) fluctuations in $\bar{\Delta}$. One may summarize the group structure for the dynamical degrees of freedom for each regime through the map
\begin{eqnarray} 
\underbrace{ \;\; p, \, q_\beta  \gg q_\Delta  - \textrm{Elementary Fermions} \;\;}_\text{$U(1)_N  \times SL(2, \mathbb{R})_{\mathrm{boost}}$ }  \;\;   \longleftrightarrow   \;\;    \underbrace{ \;\; p   , \, q_\beta   \ll q_\Delta   - \textrm{SCS} \;\;}_\text{$SL(2, \mathbb{C})_{\Delta - \mathrm{boost}} \times SU(2)_\beta \times  U(2)_\mathrm{charge}$ }   \, . 
   \end{eqnarray}

\subsection{Difermion current as gauge potential}

Let us turn our attention to the phase of $\bar{\Delta}$ to show that it gives rise to a gauge structure. For clarity, we first focus on the fermion kinetic, mass, and chemical potential terms in the Lagrange density $\mathcal{L}_\psi$. Expanded, these give
\begin{eqnarray}
\hspace{-1pc} \mathcal{L}_\psi  &=& i \psi_1^* \bar{\partial}  \,  \psi_1  +    i \psi_2^* \partial  \,  \psi_2    - m \left(  \psi_1^* \psi_2 +   \psi_2^*  \psi_1 \right)  +  \mu \left( |\psi_1|^2 + |\psi_2|^2      \right)                               \, . 
\end{eqnarray}
Let us now introduce the spin-charge decomposition in Eq.~(\ref{basictrans}) and take a general approach for breaking the internal $U(1) \leq SU(2)$ symmetry by making the substitutions
\begin{eqnarray}
\phi_{1, 2}    &\to& e^{ i \alpha_{1, 2}} \,  \phi_{1,2} \\
e^{\pm \zeta/2} &\to&  e^{\pm  i \beta}  \, e^{\pm \zeta/2} \,, 
\end{eqnarray}
where $\phi_{1, 2}$ are quasiparticle spin components and where $\alpha_{1, 2}$, $\beta$ and $\zeta$ contain the background. This decomposition gives the spin-charge separated Lagrange density
\begin{eqnarray}
     \mathcal{L}_\psi \to   \mathcal{L}_{scs}&=&        i  \,  \phi_1^* \bar{\partial}  \,  \phi_1     +  i   \, \phi_2^* \partial  \,  \phi_-       + |\phi_1|^2    \, \bar{\partial} ( \alpha_1 -   \beta   )        +     |\phi_2|^2   \, \partial (\alpha_2  +   \beta )                                             \nonumber   \\
     &-& m \left(  \phi_1^* \phi_2  e^{-i(2 \beta + \alpha_1 -   \alpha_2) }   +   \phi_2^*  \phi_1 e^{i(2 \beta + \alpha_1 -  \alpha_2) } \right)  +       \mu \left( |\phi_1|^2     + |\phi_2|^2     \right)                                \, , 
\end{eqnarray}
where we use the short hand notation $\partial \equiv \partial_t + \partial_x$, $\bar{\partial } \equiv \partial_t - \partial_x$ and have reabsorbed the $\zeta$ factors in order to isolate the internal phase. Thus, allowing for the phases to vary independently and locally gives a model with explicit symmetry breaking at the classical level. Then taking $\alpha_1 = \alpha_2 = \alpha$ to eliminate redundancy, we can identify $\alpha = \theta_N$ and $\beta = \beta_\Delta$ (the background fermion and difermion phase, respectively), which leads to the form 
\begin{eqnarray}
\hspace{-1pc} \mathcal{L}_{scs}  &=&     i  \, \phi_1^* \bar{\partial}  \,  \phi_1   +   i   \, \phi_2^* \partial  \,  \phi_2      +                          |\phi_1|^2    \, \bar{\partial} ( \theta_N -  \beta_\Delta   )        +    |\phi_2|^2   \, \partial (\theta_N  +  \beta_\Delta )      \nonumber \\
       &-& m \left(  \phi_1^* \phi_2  e^{-i2 \beta_\Delta  }   +   \phi_2^*  \phi_1 e^{i 2 \beta_\Delta  } \right)  +       \mu \left( |\phi_1|^2     + |\phi_2|^2          \right)                                \, .  \label{quasidesc}
\end{eqnarray}
Local gauge invariance can only be recovered if we now introduce a gauge field into the original Lagrange density as covariant derivative that absorbs the local phase fluctuations, i.e., 
\begin{eqnarray}
\hspace{-1pc} \mathcal{L}_0  \; \to \;  i  \bar{\psi} \, \gamma^\mu \! \left( \partial_\mu   - i \mathcal{A}_\mu \right) \psi    \, ,  
\end{eqnarray}
with the components of the gauge field transforming as 
\begin{eqnarray}
\mathcal{A}_0  &\to& \mathcal{A}_0   -   \left(  \partial_t  \theta_N  +  \partial_x  \beta_\Delta \right)       \, , \\
\mathcal{A}_1 &\to&   \mathcal{A}_1      -     \left(  \partial_x \theta_N  +  \partial_t \beta_\Delta \right) \,      \, , 
\end{eqnarray}
for unit charge. 

As a few examples of gauge fixing, consider that, for pure gauge, the Lorentz gauge is realized if $\partial_\mu \mathcal{A}^\mu = 0 \; \Rightarrow \;  \partial_t \mathcal{A}_0 - \partial_x \mathcal{A}_1 =    \bar{\partial} \partial \theta_N  = 0$, the Coulomb gauge if $\partial_x \mathcal{A}_1 = 0 \; \Rightarrow \; \partial_x^2 \theta_N + \partial_x \partial_t \beta_\Delta = 0$, the Weyl gauge if $\mathcal{A}_0  = 0 \; \Rightarrow \; \partial_t \theta_N + \partial_x  \beta_\Delta = 0$, and in the presence of a source the Dirac gauge is equivalent to $\mathcal{A}_\mu \mathcal{A}^\mu = k^2$. Thus, gauge fixing in the present context reflects a particular mixing between the phase associated with fermion number and that of the difermion field. In any case, the field strength tensor can be computed generally and found to be
\begin{eqnarray}
\mathcal{F}^{0 ,  1} &=& \partial_t  \mathcal{A}_1 - \partial_x  \mathcal{A}_0 = -  \left( \partial_t^2 -  \partial_x^2 \right) \beta_\Delta    = - \mathcal{E}   \, ,  \label{efield} \\
\mathcal{F}^{1, 0} &=& - \mathcal{F}^{0, 1}  \,, \;\; \mathcal{F}^{0,0} = \mathcal{F}^{1, 1} =0 \, , 
\end{eqnarray}
from which one sees that the emergent \emph{electric field} $\mathcal{E}$ that binds spin and charge degrees of freedom is generated by fluctuations in the difermion phase $\beta_\Delta$, but not the overall number phase $\theta_N$. Moreover, we see that the electric field vanishes when the difermion phase completely decouples from the density. It is important to point out that the electric field as defined in our context is generated by the relative phase between spin components which is zero for vacuum spinor states. It is a feature of dressed states associated purely with the background.

\section{Regimes}

In this section we examine effects of the gauge field and how it delineates fundamental versus quasiparticle regimes; spin and charge bound in the former and free to propagate independently in the latter. First, it is interesting to note that the gauge field $\mathcal{A}_\mu$ can be absorbed into a modified chemical potential that includes a chiral term which can be accounted for through a spin imbalance:
\begin{eqnarray}
\mu_1  =   \mu -  (\mathcal{A}_0 -  \mathcal{A}_1 ) \, , \;\;  \;\; \mu_2  =  \mu -  (\mathcal{A}_0 +  \mathcal{A}_1 ) \, . 
\end{eqnarray}
These results can then be combined into effective chiral and fermion number chemical potentials
\begin{eqnarray}
\mu_5 \equiv \frac{\mu_1 - \mu_2}{2}   =  \mathcal{A}_1 =   - (  \partial_t  \beta_\Delta  +   \partial_x  \theta_N)
               \, , \;\;  \;\;     \bar{\mu} \equiv   \frac{ \mu_1 + \mu_2 }{2}  =  \mu -   \mathcal{A}_0   =    \mu  +    \partial_t  \theta_N  +  \partial_x  \beta_\Delta   \, .  \label{spinpol}
\end{eqnarray}
These expressions give us some insight into the effect of the difermion phase $\beta_\Delta$. For Fourier modes, we can identify the difermion frequency and wavenumber to be $\omega_\Delta  =  - \partial_t \beta_\Delta$, $k_\Delta =  \partial_x \beta_\Delta$, and the fermion number frequency and wavenumber $\omega_N  =  - \partial_t \theta_N$, $k_N = \partial_x \theta_N$, so that 
\begin{eqnarray}
\mu_5 = \omega_\Delta - k_N
               \, \;\; \mathrm{and} \;\;  \;\;     \bar{\mu}  =   \mu -    \omega_N   +  k_\Delta  \, .  \label{spinpol2}
\end{eqnarray}

Several regimes can be identified from these:

\begin{enumerate}

\item    \emph{In-vacuum fermions}.   At large fermion momentum ($k_N \gg m$), vanishing chemical potential ($\omega_N \simeq k_N$) and difermion field, Eq.~(\ref{spinpol2}) becomes $\mu_5 \simeq  \bar{\mu} \simeq -  k_N$. Thus, in the vacuum $\omega_N$ is the single-particle energy proportional to the momentum $k_N$ (factors of $\hbar$ aside) and boosts couple to spin through the chiral imbalance induced by $\mu_5$. In the absence of the difermion field, chiral imbalance is induced purely by the fermion momentum. This is the usual spin-momentum coupling for $(1+1)$d relativistic fermions where one spin component increases and the other decreases with increasing momentum.

   \item   \emph{In-medium fermions with manifest $U(1)_\Delta$ symmetry}.    Here we consider a mixed phase with finite $\mu$ and $m$ but small nonlinearity in the scalar field equation that governs the difermion field. This would correspond to the quantum regime where $\langle \Delta \rangle =  0$ and depletion of the difermion condensate is large. In this regime, stiffness of the difermion phase $\beta_\Delta$ vanishes leading to large quantum fluctuations in this field. From a mean-field perspective, $\langle \omega_\Delta \rangle$, $\langle k_\Delta \rangle \simeq 0$ so that $\mu_5 \simeq - k_N$, $\bar{\mu} \simeq \mu -  \omega_N$. Similar to the in-vacuum case, fermion spin and charge are strongly coupled in this regime, with the electric field $\mathcal{E}$ diverging due to the unregulated quantum fluctuations of the difermion phase field in the ultraviolet.

   \item     \emph{Low-energy fermionic fluctuations in broken $U(1)_\Delta$ symmetry}.     Here we consider the classical regime where $\langle \Delta \rangle \ne  0$ where all fluctuations are small with respect to a large difermion condensate. We will show that in this regime the difermion density and phase decouple such that $\omega_\Delta \sim k_\Delta$ but where $\omega_N \ne k_N$, since the fermion frequency encodes information about the density $\rho_\Delta$ but the wavenumber does not (which we show below). The chemical potentials can then be expressed as $\mu_5 = k_\Delta - k_N$ and $\bar{\mu} = \mu + k_\Delta - \omega_N$, which displays the chemical potentials as two independent degrees of freedom. The characterization of Eq.~(\ref{quasidesc}) as the quasiparticle description for spin-charge separation is justified as the overall number and spin densities are determined independently through $\bar{\mu}$ and $\mu_5$. Thus, fermionic fluctuations in this regime are spinor quasiparticles whose spin and charge are driven by the decoupling of density and phase fluctuations in the underlying difermion field.

\item     \emph{High-energy fermionic fluctuations in broken $U(1)_\Delta$ symmetry}.   This regime addresses the limit as the free fermion momentum approaches that of the bound fermions that comprise the difermion field. For excitations in the ultraviolet, the intermediate steps when deriving Eq.~(\ref{efield}), where we assume the cancellation $\partial_x \partial_t \theta_N - \partial_t \partial_x \theta_N = 0$, should be modified if one takes the time derivative of the phase to be the single-particle energy. From this perspective, the commutator of mixed partial derivatives should be replaced by $[ \partial_x , \, \hat{h}]$, where $\hat{h}$ is the fermion single-particle Hamiltonian. This introduces a term proportional to the background potential felt by the free fermions which supplies an additional contribution the electric field. Thus, even with vanishing $\partial_\mu \beta_\Delta   \sim 0$, we have $\mathcal{E} = \left(  \partial_x u_\mathrm{bg} \right)\theta_N$. If we assume a uniform difermion density, $u_\mathrm{bg}$ must be associated with purely fermionic fluctuations at the molecular scale of the difermion. The associated charge density is then given by $\rho_\mathcal{E} = \partial_x \mathcal{E} =     \left(  \partial_x^2  u_\mathrm{bg} \right)\theta_N +  \left(  \partial_x  u_\mathrm{bg} \right) \left( \partial_x \theta_N  \right)$. For slowly varying $\theta_N$ (small $p_F$) one is free to choose $\theta_N \simeq 0$, in which case $\rho_\mathcal{E} \to 0$. It is only when $\theta_N$ varies considerably over the size of the difermion molecule, measured by $\partial_xu_\mathrm{bg}$ and $\partial_x^2u_\mathrm{bg}$, that we begin to see large values of $\rho_\mathcal{E}$.

\end{enumerate}

In order to better understand how SCS is realized it helps to look at solutions of the underlying equations for the difermion field and examine regions in which the phase and density currents decouple. This coincides with vanishing $\mathcal{E}$ in Eq.~(\ref{efield}) which signals the onset of SCS. Returning to the classical equations of motion for the difermion density and phase in Eq.~(\ref{densityphasedec}), along with Eq.~(\ref{efield}) and $U(1)_\Delta$ broken symmetry, we have
\begin{eqnarray}
\partial_\mu \partial^\mu \rho_\Delta - m_\Delta^2 \rho_\Delta + \frac{g_\Delta}{6} \rho_\Delta^3  &=&  \rho_\Delta \left[ (   \partial_t \beta_\Delta )^2 -  (\partial_x \beta_\Delta)^2 \right]  \, \label{densityeq} ,\\
\partial_t \left( \rho_\Delta \partial_t \beta_\Delta    \right)  &=&    \partial_x \left(  \rho_\Delta \partial_x \beta_\Delta \right)\,  , \\
\mathcal{E}    &=&  \left(  \partial_t^2    -  \partial_x^2 \right) \beta_\Delta   =     \left(  \partial_t    -  \partial_x \right)    \left(  \partial_t   +  \partial_x \right)           \beta_\Delta     \, . \label{FieldDef}
\end{eqnarray}
The second and third equations can be integrated to give
\begin{eqnarray}
\beta_\Delta' =  \frac{C}{\rho_\Delta}  \;\; , \;\;\;\; \mathcal{E} = C(k^2 \pm  \omega^2 ) \frac{\rho_\Delta' }{\rho_\Delta^2} \, ,  \label{E}
\end{eqnarray}
where we have assumed the forms $\beta_\Delta( kx \pm \omega t )$, $\rho_\Delta( kx \pm \omega t)$, and $C$ is an integration constant. Using these, we can decouple Eq.~(\ref{densityeq}) to give
\begin{eqnarray}
\rho_\Delta \partial_\mu \partial^\mu \rho_\Delta - m_\Delta^2 \rho_\Delta^2 + \frac{g_\Delta}{6} \rho_\Delta^4  &=&  C ( \omega^2 - k^2 ) \, , 
\end{eqnarray}
which has standard kink solutions when $\omega = \pm k$:
\begin{eqnarray}
\rho_\Delta(x, t) = \frac{m_\Delta }{\sqrt{g_\Delta/3}}  \mathrm{tanh} \left[ m_\Delta (x \pm    t )\right]  \, . \label{Ceq}
\end{eqnarray}
Soliton solutions are also possible for the more general case $\omega \ne \pm k$. These kinks interpolate between the four possible asymptotic values
\begin{eqnarray}
\lim_{x \to \pm \infty} \rho_\Delta(x)  \sim   \pm   \frac{m_\Delta}{\sqrt{g_\Delta /3}}   \,  \left[  1 \pm   \sqrt{ 1 + \frac{2 g_\Delta}{3 m_\Delta^4} C ( \omega^2 - k^2 )}      \;      \right]^{1/2}    \, , \label{asymptotic}
\end{eqnarray}
with behavior near the core given by
\begin{eqnarray}
\lim_{x  \to 0} \rho_\Delta( x) \sim C   e^{ - \mathrm{erf}^{-1} \left[  \pm i \sqrt{\frac{2}{\pi}}   (C_1  + x)       \right]^2                    } \, . 
\end{eqnarray}
Such kink solitons interpolate between the distinct spin-charge regimes enumerated above:

\begin{enumerate}[label=(\roman*)]

\item \emph{Core region}. Here, the difermion density goes to zero so that the stiffness associated with bosonic phase fields vanishes and the classical solution for $\beta_\Delta$ diverges. The expression for $\mathcal{E}$ in Eq.~(\ref{E}) also diverge, in general, but still contains the special case where the density and phase decouple ($k = \pm  \omega$). In the extreme quantum regime, these modes have zero weighted measure in the vast landscape of energetically favorable quantum states. Rather, the dominant modes are large nonuniform and time-dependent fluctuations for which $k \ne \pm \omega$. Thus, this region is characterized by arbitrarily large fluctuations $\delta \mathcal{E} = \left( \partial_t^2 -  \partial_x^2 \right) \delta \beta_\Delta$ with unbounded measure. In short, this regime is more aptly characterized through the \emph{dual} description of strongly bound spin and charge, i.e., vacuum fermions, thus the first regime discussed above.

\item \emph{Tail regions}. These are the four possible asymptotic regions far from the soliton core where the difermion condensate is uniform and large such that $\langle \rho_\Delta \rangle \gg C$, in Eq.~(\ref{Ceq}). Fluctuations here are in the infrared with vanishing $\mathcal{E}$ for which the quasiparticle description in terms of independent spin and charge degrees of freedom is appropriate. This is consistent with the third regime above.

\end{enumerate}

\section{Quasiparticle solutions and spin-charge theorems }

In this section, we find explicit quasiparticle solutions and use these to motivate conclusions regarding a broader class of interactions. To proceed, we first assume fluctuations over a uniform constant difermion background. Linearizing the equations of motion for the difermion field by taking $\rho_\Delta \to \rho_0 + \rho_\Delta$, where we assume small density fluctuations $\rho_\Delta$ over the uniform background $\rho_0$, such that $\rho_\Delta \ll  \rho_0$, we obtain
\begin{eqnarray}
\partial_\mu \partial^\mu \rho_\Delta +    \left(  m_\Delta^2  +    \frac{g_\Delta}{2} \rho_0^2 \right)  \rho_\Delta  &=& 0 \, ,  \label{Rhoden2} \\
(\partial_t \pm  \partial_x ) \beta_\Delta &=& 0 \, . \label{Betaphase2} 
\end{eqnarray}
Solving the first equation by assuming the form $\rho_\Delta(k_\rho  x \pm  \omega_\rho t) \equiv \rho_\Delta(u)$ yields plane wave solution (when $\omega_\rho^2 > k_\rho^2$) 
\begin{eqnarray}
 \rho_\Delta''(u)    &=&   - \frac{  m_\Delta^2  +   g_\Delta \rho_0^2 /2     }{(\omega_\rho^2 \pm  k_\rho^2 )}    \rho_\Delta(u) \\
 \Rightarrow \;\;\;   \;\;\; \rho_\Delta(k_\rho x \pm  \omega_\rho t) &=& A_\rho \,  \mathrm{cos} \left[ \frac{  m_\Delta^2  +   g_\Delta \rho_0^2 /2     }{(\omega_\rho^2 \pm  k_\rho^2 )} (k_\rho x \pm  \omega_\rho t) \right] +         B_\rho \,  \mathrm{sin} \left[ \frac{  m_\Delta^2  +   g_\Delta  \rho_0^2 /2     }{(\omega_\rho^2 \pm  k_\rho^2 )} (k_\rho x \pm  \omega_\rho t) \right] \,. \label{Rhosoln} \end{eqnarray}
Similarly, the equation for the phase $\beta_\Delta$ yields
\begin{eqnarray}
\beta_\Delta( x ,  t) &=& f( k_\beta(x \pm   t) )  +  \textrm{constant} \;\; . \label{Betasoln}
\end{eqnarray}
To reintroduce the emergent electric field $\mathcal{E}$, we must express the continuity equation in Eq.~(\ref{densityphasedec}) to lowest order in the density and phase fluctuations and use Eq.~(\ref{efield}), which gives the expression
\begin{eqnarray}
 \mathcal{E} = \partial_\mu   \mathrm{ln}  \left(  1 + \rho_\Delta/\rho_0 \right)^{-1}  \partial^\mu  \beta_\Delta    \,   . 
 \end{eqnarray}
Hence, $\mathcal{E}$ is a measure of the coupling between the difermion density and phase. It is important to note that this result is valid not only for vanishingly small but also for large fluctuations in the fields and assumes only a uniform background $\rho_0$. Specifically, for a large values of the background $\rho_0$ (large condensate), both density and phase fluctuations are suppressed and we obtain the asymptotic behavior for the electric field 
\begin{eqnarray}
\mathrm{lim}_{\rho_\Delta/\rho_0 \to \, 0} \;  \mathcal{E} \sim  - \frac{1}{\rho_0} \,  \partial_\mu \rho_\Delta   \partial^\mu \beta_\Delta \to 0 \, , 
\end{eqnarray}
where $\mathcal{E}$ is suppressed by second order in the fluctuations.

In the limit of weak nonlinearity where both the difermion condensate is small and most of the difermions have dissociated, i.e., $\langle \rho_\Delta  \rangle , \, \rho_\Delta  \to 0$, and assuming traveling wave forms, Eqs.~(\ref{densityeq})-(\ref{FieldDef}) reduce to 
\begin{eqnarray}
 \rho_\Delta'  &=&      \left[     \frac{m_\Delta^2 }{\omega_\rho^2 - k_\rho^2 }\,   \rho_\Delta^2  \,  +     \,C^2  \left(  1 -      2   \frac{ \omega_\beta \omega_\rho - k_\beta k_\rho  }{    \omega_\beta^2 - k_\beta^2 }   \right)^{-1}     \, \left(    \frac{ \omega_\beta^2 - k_\beta^2 }{    \omega_\rho^2 - k_\rho^2 }  \right)   \rho_\Delta^{2   -   4      \frac{ \omega_\beta \omega_\rho - k_\beta k_\rho  }{    \omega_\beta^2 - k_\beta^2 }    }   \right]^{1/2}   \, \label{densityeq2} ,\\
   \beta_\Delta'      &=&  C  \left( \frac{1}{ \rho_\Delta}  \right)^{  2 \frac{ \omega_\beta \omega_\rho - k_\beta k_\rho  }{    \omega_\beta^2 - k_\beta^2 }      }   \,  , \\
   \mathcal{E}    &=&    - \left(  \omega_\rho \omega_\beta  - k_\rho k_\beta   \right)  \frac{\rho_\Delta'    \beta_\Delta'}{  \rho_\Delta}    \, .  \label{FieldDef2}
\end{eqnarray}
 when $\omega_\beta \ne  \omega_\rho$, $k_\beta \ne k_\rho$, and 
 \begin{eqnarray}
 \rho_\Delta'  &=&      \left[     \frac{m_\Delta^2 }{\omega_\rho^2 - k_\rho^2 }\,   \rho_\Delta^2  \,  +     \,  C^2   \ln(\rho_\Delta^2)   \right]^{1/2}   \, \label{densityeq2} ,\\
   \beta_\Delta'      &=&  C  \frac{1}{ \rho_\Delta }    \,  , \\
   \mathcal{E}    &=&    - 2 \left(  \omega_\rho^2   - k_\rho^2  \right)  \frac{\rho_\Delta'    \beta_\Delta'}{  \rho_\Delta}    \, ,   \label{FieldDef2}
\end{eqnarray} 
for the special case $\omega_\beta = 2  \omega_\rho$ and $k_\beta = 2  k_\rho$. In either case, divergent solutions appear indicating breakdown of the classical equations when $\rho_\Delta \to 0$. Thus, classically allowed solutions are possible when 
\begin{eqnarray}
 \frac{1}{2}  >   \frac{ \omega_\beta \omega_\rho - k_\beta k_\rho  }{    \omega_\beta^2 - k_\beta^2 }  > 0  \, . 
\end{eqnarray}
We are now in position to make a general statement regarding spin-charge separation for paired Dirac fermions.
\vspace{1pc}

\noindent{\bf Theorem 1.1}    \emph{Let $\Delta(x, t): \mathbb{R}^{(1,1)}  \to \mathbb{C}^1$ be a bilinear spinor pairing field where $\Delta$ acts as a scalar potential in the spinor theory. Let $\Delta$ solve a complex scalar wave equation with repulsive cubic nonlinearity. In the strong nonlinear regime for $\Delta$, states in the spinor theory smoothly connect to the zero-momentum frame through a product of independent Lorentz transformations of the form $\psi_i(x)   \to   S\left[  \Lambda_i'     \right]_i^i \psi_i \hspace{-.2pc}\left( \Lambda^{-1}_i  x \right)$, where $S\left[ \Lambda \right] \in SL\left(2, \mathbb{R} \right)$ and $\Lambda_i , \Lambda_i^\prime \in O\left( 1, 1 \right)$ are different spin-indexed coordinate Lorentz transformations. Moreover, In this limit, spinor components decouple and provide a representation of spin-charge separation. In the limit of weak nonlinearity in the pairing field, fermions revert to their vacuum form and transform in the usual way under the Lorentz group.   }

\vspace{1pc}
\noindent \emph{Proof}. The task is to incorporate solutions Eqs.~(\ref{Rhosoln})-(\ref{Betasoln}) as external potentials in the spinor equation for $\psi$. Applying the decompositions $\psi_{1, 2} = \rho_{1, 2} \, e^{i \phi_{1, 2}}$, $\Delta = \rho_\Delta e^{i \beta_\Delta}$, the equations of motion for the real and imaginary parts of the spinor components decouple into
\begin{eqnarray}
( \partial_t - \partial_x ) \rho_1  +  \mathrm{cos}\beta_\Delta\,   \rho_\Delta \rho_1    + m \rho_2  \, \mathrm{sin}(\phi_2 - \phi_1 )  &=& 0 \, , \label{decomp1} \\
( \partial_t - \partial_x ) \phi_1  - \mathrm{sin}\beta_\Delta \,  \rho_\Delta  - m  (\rho_2/\rho_1)  \,       \mathrm{cos}(\phi_2 - \phi_1 )   + \mu    &=& 0 \, , \\
( \partial_t + \partial_x ) \rho_2  - \mathrm{cos} \beta_\Delta \,  \rho_\Delta \rho_2      -    m \rho_1  \, \mathrm{sin}(\phi_2 - \phi_1 )                &=& 0 \, , \\
( \partial_t + \partial_x ) \phi_2  +  \mathrm{sin}\beta_\Delta \,   \rho_\Delta     - m  (\rho_1/\rho_2)  \,       \mathrm{cos}(\phi_2 - \phi_1 )   + \mu     &=& 0    \label{decomp4}\, . 
\end{eqnarray}
In the strong nonlinear regime for $\Delta$ (large material limit where $\mu, \, m \to 0$), these yield the amplitude solutions 
\begin{eqnarray}
\rho_1  =   c_1  \, e^{  (\omega_{\rho_1} / k_{\rho_1} -1 )^{-1} \!  \int  \! dx  \,   \mathrm{cos} \beta_\Delta \,  \rho_\Delta } \; , \;\;\;   \;\;\;  \rho_2  =   c_2  \, e^{ -  (\omega_{\rho_2} / k_{\rho_2} +1 )^{-1}  \!  \int  \! dx  \,   \mathrm{cos} \beta_\Delta \,  \rho_\Delta } \, , \label{rhosolns}
\end{eqnarray}
and phase solutions 
\begin{eqnarray}
\phi_1  =     (\omega_{\phi_1} / k_{\phi_1} -1 )^{-1} \!  \int  \! dx  \,   \mathrm{sin} \beta_\Delta \,  \rho_\Delta  \; , \;\;\; \;\;\; \phi_2   =    -  (\omega_{\phi_2} / k_{\phi_2} +   1 )^{-1} \!  \int  \! dx  \,   \mathrm{sin} \beta_\Delta \,  \rho_\Delta \, . \label{phisolns}
\end{eqnarray}
For strong nonlinearity $k_\beta  \to 0$, and to zeroth order in $k_\beta$ we have $\beta_\Delta \simeq \mathrm{constant}$, in which case spinor solutions take the form 
 \begin{eqnarray}
\psi(x, t)  &=& \left( \begin{array}{ll}
  c_1 \,    e^{  i   k_{\phi_1} (x + t)  +    (\omega_{\rho_1} / k_{\rho_1} -1 )^{-1}   ( \kappa k_\rho )^{-1}  \left\{   A_\rho  \, \mathrm{sin}\left[ \kappa ( k_\rho x - \omega_\rho t )  \right]      -  B_\rho  \, \mathrm{cos}\left[ \kappa ( k_\rho x - \omega_\rho t )  \right]      \right\}           }                         \\
   c_2  \,     e^{  i   k_{\phi_2} (x - t)  -      (\omega_{\rho_2} / k_{\rho_2}  + 1 )^{-1}   ( \kappa k_\rho )^{-1}  \left\{   A _\rho \, \mathrm{sin}\left[ \kappa ( k_\rho x - \omega_\rho t )  \right]      -  B_\rho  \, \mathrm{cos}\left[ \kappa ( k_\rho x - \omega_\rho t )  \right]      \right\}           }  
                  \end{array} \right) \, , \label{psisolution}
\end{eqnarray} 
where $\kappa \equiv  ( m_\Delta^2  +   g_\Delta \rho_0^2 /2  ) (\omega_\rho^2 - k_\rho^2 )^{-1}$. From these, we can deduce the first-order expressions 
\begin{eqnarray}
\beta_\Delta &=& \frac{1}{2} \left(  \phi_1 - \phi_2    \right)  = \frac{1}{2} \left(  k_{\phi_1} - k_{\phi_2} \right) \left( x + t \right)  + C_\beta \; ,\\
\theta_N  &=&    \frac{1}{2} \left(  \phi_1 +  \phi_2    \right)   =    \frac{1}{2} \left(  k_{\phi_1} + k_{\phi_2} \right) \left[   x +    \left(  1  + 2   \frac{ \sin( C_\beta) A_\rho }{k_{\phi_1}+ k_{\phi_2} }      \right)  t \right]   \, , \\
\mathcal{E}    &=& \left(\partial_t^2 - \partial_x^2 \right)  \beta_\Delta \,   \simeq \,   0 \, .
\end{eqnarray}
The coefficients $A^{i}$ and $B^{i}$, which enfold the frequency factors into $A_\rho$ and $B_\rho$, must transform as sums of spinor component bilinears since $\rho_\Delta = |\Delta|$. Under an infinitesimal Lorentz transformation $\Lambda_\epsilon$ we then have 
\begin{eqnarray}
A^i    \, \sim   \,     A_{a, b}^i  \,  \psi^a \psi^b         &\to&   A_{a, b}^i   \,   S[ \Lambda_\epsilon  ]^a_c  \, \psi^c    \, S[ \Lambda_\epsilon  ]^b_d  \, \psi^d   \,  \simeq  \,    A_{a, b}^i \,  ( \delta_c^a  + g_c^a    \epsilon )  \, ( \delta_d^b  + g_d^b    \epsilon )  \,  \psi^c  \,  \psi^d  + \dots \\
    &\simeq&      A_{a, b}^i  \,  \psi^a \psi^b \, + \, \left(    A_{a, b}^i  \,  \delta_c^a   \,  \psi^c    g_d^b  \,  \psi^d      +                A_{a, b}^i is  \,  \delta_d^b   \,  \psi^d    g_c^a  \,  \psi^c                \right) \epsilon + \dots   \\
    &\equiv&    A^i   \,  +  \, g_i^i  \, \epsilon \, + \, \dots \, , 
    \end{eqnarray} 
 and similarly for $B^i$. In the absence of propagation, the components in Eq.~(\ref{psisolution}) acquire the pure spinor form transforming as 
 \begin{eqnarray}
 c_i \,  \to  \, e^{ (-1)^i  g_i^i \,  \epsilon }  \,  c_i  \,   =     \,   S\left[  \Lambda_\epsilon^i        \right]_i^i   c_i    \;  , 
\end{eqnarray}
which, in the presence of space and time dependence, enlarges to
\begin{eqnarray}
 \psi_i(x) \,  \to       \,   S\left[  \Lambda_\epsilon^i        \right]_i^i   \psi_i( \Lambda^{-1} x )  \, .  
 \end{eqnarray}

 In contrast, in the weak nonlinear limit for $\Delta$, $\rho_\Delta \to 0$, and taking $\phi_2 - \phi_1 = \pi$ while retaining $\mu , \, m \ne 0$, Eqs.~(\ref{decomp1})-(\ref{decomp4}) reduce to the vacuum spinor equations. Combining Eq.~(90) and Eq.~(92), we obtain 
\begin{eqnarray}
  \rho_\Delta  &=&      \left[  C \left(  2      \frac{ \omega_\beta \omega_\rho - k_\beta k_\rho  }{    \omega_\beta^2 - k_\beta^2 } \right)       \left(  1 -      2   \frac{ \omega_\beta \omega_\rho - k_\beta k_\rho  }{    \omega_\beta^2 - k_\beta^2 }   \right)^{-1/2}     \, \left(    \frac{ \omega_\beta^2 - k_\beta^2 }{    \omega_\rho^2 - k_\rho^2 }  \right)^{1/2}       \left(  k_\rho x - \omega_\rho t     \right)                             \right]^{      \frac{     \omega_\beta^2 - k_\beta^2 } {2   ( \omega_\beta \omega_\rho - k_\beta k_\rho ) }    }    \,  ,\\
   \beta_\Delta       &=&  \left[   \left(  2      \frac{ \omega_\beta \omega_\rho - k_\beta k_\rho  }{    \omega_\beta^2 - k_\beta^2 } \right)       \left(  1 -      2   \frac{ \omega_\beta \omega_\rho - k_\beta k_\rho  }{    \omega_\beta^2 - k_\beta^2 }   \right)^{-1/2}     \, \left(    \frac{ \omega_\beta^2 - k_\beta^2 }{    \omega_\rho^2 - k_\rho^2 }  \right)^{1/2}                               \right]^{-1}     \ln\left(  k_\beta  x - \omega_\beta t     \right)       \,  , \\
\mathcal{E}  &=& \left(\partial_t^2 - \partial_x^2 \right) \frac{1}{2} \left(  \phi_1 - \phi_2    \right)  = k_\beta \beta_\Delta' \cos( \beta_\Delta )   \,    \rho_\Delta         +               \sin( \beta_\Delta )    \,   \rho_\Delta'  \, \simeq \,  0 \, .
\end{eqnarray}
The electric field $\mathcal{E}$ associated with the internal $U(1)$ symmetry vanishes in both limits: due to decoupling of density and phase of the pairing field, in the case of large $\rho_\Delta$, and extreme coupling of these in the case of vanishingly small $\rho_\Delta$. In the later case, vanishing $\mathcal{E}$ arrises due to tightly bound spin and charge into $U(1)$ neutral states with details of their internal interactions encoded in the diverging oscillations of $\cos( \beta_\Delta )$ and $\sin( \beta_\Delta )$. Finally, it is interesting to note that mean-field fluctuations in $\beta_\Delta$ are inherently $\mathcal{P} \mathcal{T}$-symmetry breaking as explained in Sec.~\ref{groupstructure}.

\vspace{1pc}
\noindent{\bf Theorem 1.2}   \emph{The formulation of paired fermions in terms of independent spin and charge degrees of freedom describes a 3-dimensional cobordism $(W; M, N)$, where $M \cong S^1  \times G_M$ and $N \cong S^1  \times S^1 \times G_N$ are the vacuum and in-medium ground state manifolds, respectively, the circles $S^1$ are generic (model independent) and $G_{M, N}$ are additional model-specific groups structures.  }

\vspace{1pc}
\noindent \emph{Proof}. Condensation in the difermion field defines a natural unit interval $\sqrt{g_\Delta}\,  \langle \rho_\Delta \rangle/\sqrt{3} m_\Delta \in \mathrm{I} =  \left[ 0 , \, 1 \right]$ that interpolates between manifest and broken $U(1)_\Delta$ symmetry, parametrized by $\beta_\Delta \in [ 0, \, 2 \pi ]$ through the complex phase $(\sqrt{g_\Delta}\,  \langle \rho_\Delta \rangle/\sqrt{3} m_\Delta  )  \,   e^{ i \beta_\Delta} \cong  S_\Delta^1(I)$. Also, define $S^1_N \cong U(1)_N$. We can then identify a 3-dimensional in-medium manifold $W \cong S^1_N \times S^1_\Delta(I) \times I$ with boundary embeddings $i: M \hookrightarrow \partial W$, $j: N \hookrightarrow \partial W$ and where $M \cong S^1_N \times S^1_\Delta(1) \times \left\{ 1 \right\}$, $N \cong S^1_N \times S^1_\Delta(0) \times \left\{ 0 \right\}$, such that the boundary of $W$ is the disjoint union $\partial W = i(M) \sqcup j(N)$.

\section{conclusion}

We have shown that spin and charge decouple into distinct excitations in Dirac systems with four-fermion interactions that include a dynamical complex difermion channel. The complex pairing field introduces potential couplings into the Dirac equation that suggest a reformulation in terms of factorized dressed spinor wavefunctions. The additional factors encapsulate effects coming from the background. In particular, the density of the complex field introduces a real factor that acts as an element of $SL(2, \mathbb{R})$, an effective boost from the dressed spinor perspective. The complex phase introduces a factor that acts as an element of $U(1)$ internal to the dressed spinor, its derivates identified with a four-vector gauge potential that mediates SCS. These additional degrees of freedom effectively act by enlarging the symmetry group of fermionic sector. 

Two distinct ground states emerge through this method, connected through a crossover region: 1) Dirac vacuum -- here the density of the complex pairing field vanishes and its phase fluctuations diverge (manifest symmetry) leading to tightly bound spin and charge with spinor wave functions identified with Dirac scattering states; 2) Spin-charge vacuum -- condensation in the scalar field is large (broken symmetry) and its density and phase fluctuations decouple resulting in dressed spinors with two phase (charge) fields and a spin (boost) field, each propagating independently at distinct speeds. We have shown that this crossover can also be traversed by tuning the momentum of elementary Dirac states relative to the strength of the gauge field and the complex pairing condensate. 

We found that the onset of SCS is signaled by breaking of $\mathcal{P}\mathcal{T}$ symmetry with a corresponding Lorentz violation relative to the Dirac vacuum. The violation of Lorentz symmetry is such that spin components decouple into upper and lower charge fields under SCS connecting to the ground state of the system through a product group $SL(2, \mathbb{R})_1 \times   SL(2, \mathbb{R})_2$ and charged under the product $U(1)_1 \times U(1)_2$. This is in stark contrast to Dirac vacuum states for which the spin components are intricately connected to Lorentz boosts transforming under only one factor of $SL(2 , \mathbb{R})$ and having a single fermion number charge associated with $U(1)_N$. From an interesting mathematical perspective, the two ground states map naturally into the boundary of a three-dimensional manifold thus identifying a cobordism between Dirac and spin-charge vacua with the complex scalar condensate as interpolating dimension.

\begin{acknowledgments}
The author would like to thank the Department of Physics at Colorado School of Mines for support during the writing of this manuscript.
\end{acknowledgments}

\bibliography{SCS_Refs}

\end{document}